\documentclass[prl,twocolumn,superscriptaddress,nofootinbib,amsmath,amssymb]{revtex4-2}

\usepackage{booktabs}
\usepackage{graphicx}
\usepackage{xcolor}
\usepackage[hidelinks]{hyperref}

\newcommand{\Ft}{\tilde F}
\newcommand{\Lstar}{\Lambda_*}
\newcommand{\GeV}{\,\mathrm{GeV}}

\begin{document}

\title{Can cosmic neutrinos induce a cosmic birefringence signal?}

\author{Mario Reig}
\affiliation{Theoretical Physics Department, CERN, 1211 Geneva 23, Switzerland}

\begin{abstract}
A coupling of photons to the cosmic neutrino asymmetry current has been proposed to explain the hinted isotropic rotation of CMB polarization. Gauge invariant UV completions require a Kalb--Ramond field that is exactly dual to a massless axion. We show that, in this case, the product of the axion-photon coupling and the axion-neutrino current coupling is fixed by the signal, $g_{a\gamma}/f_\ell=1/\Lstar^2$ with $\Lstar\simeq10$\,TeV. Using $SU(2)_L$ gauge invariance we show that the dual axion couples to electrons as $g_{a\gamma}g_{ae}=m_e/\Lstar^2\simeq5\times10^{-12}\GeV^{-1}$. This value is excluded by the measured electron magnetic moment combined with light-shining-through-a-wall searches, as well as by stellar evolution. Conserved anomaly-free currents may evade these bounds, but then either a contact interaction among SM fermions excludes the model, or the relic neutrinos decouple from the signal. We conclude that cosmic neutrinos cannot induce a birefringence signal.
\end{abstract}

\maketitle

\textit{Introduction.}---The polarization of the cosmic microwave background (CMB) may provide the first evidence of parity violation on cosmological scales. Recent analyses of Planck, WMAP and ACT data find an isotropic rotation of the plane of linear polarization by $\beta\simeq0.2$--$0.3^\circ$~\cite{Minami:2020odp,Diego-Palazuelos:2022dsq,Eskilt:2022cff,ACT:2025fju,Eskilt:2026imm}. If cosmological, this signal requires physics beyond the Standard Model (SM)~\cite{Komatsu:2022nvu}. The standard interpretation is a light axion field $\varphi$ coupled to photons as $\varphi F_{\mu\nu}\Ft^{\mu\nu}$, with $F_{\mu\nu}$ the electromagnetic field strength. $\varphi$ could play the role of dark energy or a small fraction of dark matter~\cite{Harari:1992ea,Carroll:1998zi,Lue:1998mq,Fedderke:2019ajk}, although it need not be either. The rotation can also be sourced by topological defects, such as axion strings or domain walls as photons travel through the network~\cite{Agrawal:2019lkr,Takahashi:2020tqv,Jain:2021shf,Kitajima:2022jzz,Ferreira:2023jbu,Lee:2025yvn,Kaloper:2026ygk,DAmico:2026fuy}.

An economical alternative uses the interaction with a cosmic neutrino asymmetry and seems to require no new field. If the cosmic neutrino background carries a lepton asymmetry, with net number density $\Delta n_\nu=n_\nu-n_{\bar\nu}$, its current $j^\mu=(\Delta n_\nu,\vec0)$ selects a preferred time direction. A coupling of the type $j_\mu A_\nu\Ft^{\mu\nu}$ turns it into a parity-violating medium~\cite{Geng:2007va,Ho:2010nk,Zhou:2023aqh}. %
The required neutrino asymmetry, $\xi_{\nu_e}\sim0.05$, where $\xi_{\nu_e}=\mu_{\nu_e}/T_\nu$ is the electron-neutrino degeneracy parameter, is close to the one suggested by the low primordial helium abundance measured by EMPRESS~\cite{Matsumoto:2022tlr} (see also Refs.~\cite{Escudero:2022okz,Domcke:2025lzg,Toda:2026vrf}). Recently, Yin and Komatsu have shown that the scenario predicts a distinctive shape of the $EB$ cross-power spectrum, which distinguishes it from dark-energy and dark-matter axions, turning it into a testable proposal~\cite{Yin:2026nu}. Its anisotropies have also been
computed recently~\cite{Kasai:2026lep}. 

The coupling between photons and the neutrino current required for the birefringence signal, however, is not gauge invariant. Restoring gauge invariance requires additional degrees of freedom~\cite{Nakai:2023zdr}. In this Letter we show that the completion used in the literature, a Kalb--Ramond 2-form~\cite{Geng:2007va,Ho:2010nk}, is dual to a massless axion with couplings that are not free: the product of its photon coupling and its coupling to the lepton current is fixed by the birefringence signal, independently of the unknown 2-form normalization. We show that the photon is protected from acquiring a mass, so the dual axion is the only new light state. $SU(2)_L$ gauge invariance then forces this axion to couple to electrons with a strength that laboratory and astrophysical data exclude. In this completion, neutrino-current birefringence is nothing but an axion model with couplings that experiments excludes. We also discuss currents that avoid the tree-level electron coupling, showing they are equally excluded.

\textit{The operator.}---The neutrino current explanatio rests on the operator
\begin{equation}
\mathcal L_J=-\frac12\frac{b_J}{M^2}\,j_\mu A_\nu\Ft^{\mu\nu},
\qquad j^\mu=\bar\nu_{e}\gamma^\mu P_L\nu_{e} ,
\label{eq:LJ}
\end{equation}
with $A_\mu$ the photon field, $\Ft^{\mu\nu}=\tfrac12\epsilon^{\mu\nu\rho\sigma}F_{\rho\sigma}$, $P_L=(1-\gamma_5)/2$, $M$ the scale of new physics and $b_J$ a dimensionless coefficient. On the cosmological background, this term rotates the polarization of CMB photons~\cite{Carroll:1989vb} by
\begin{equation}\label{eq:bire_angle}
    \beta=\tfrac12(b_J/M^2)\int\Delta n_{\nu_e}\,dt\,,
\end{equation}
with $t$ cosmic time and the integral running from recombination until today. The fit of Yin and Komatsu to Planck $EB$ data gives $M/\sqrt{b_J\xi_{\nu_e}}=44^{+12}_{-7}$\,TeV~\cite{Yin:2026nu}, so the relevant scale is
\begin{equation}
\Lstar\equiv\frac{M}{\sqrt{b_J}}=44\,\mathrm{TeV}\,\sqrt{\xi_{\nu_e}}\simeq10\,\mathrm{TeV}.
\end{equation}

Under an EM gauge transformation $A\to A+\partial\chi$,
\begin{equation}
\delta\mathcal L_J=\frac{b_J}{4M^2}\,\chi\,\big(\partial_\nu j_\mu-\partial_\mu j_\nu\big)\Ft^{\mu\nu}
+\text{total derivative},
\label{eq:dLJ}
\end{equation}
which vanishes for the homogeneous background but not as an operator identity. 
A vector $p_\mu$ couples to $K^\mu\equiv A_\nu\Ft^{\mu\nu}$ in a gauge invariant way only if it is a gradient, $p_\mu=\partial_\mu\theta$, and then $p_\mu K^\mu=-\tfrac12\theta F\Ft$ up to a total derivative. A local, Lorentz-invariant completion with a massless photon must therefore contain a new light
pseudoscalar playing the role of $\theta$, which the completion below makes explicit.

\textit{The Kalb--Ramond completion.}---Refs.~\cite{Geng:2007va,Ho:2010nk} introduce a 2-form $B$~\cite{Kalb:1974yc} with the gauge-invariant field strength
\begin{equation}
H_{\mu\nu\rho}=3\,\partial_{[\mu}B_{\nu\rho]}+3\,A_{[\mu}F_{\nu\rho]},
\qquad \delta A=d\chi,\quad \delta B=-\chi F,
\end{equation}
where $[\dots]$ denotes antisymmetrization. Its 4d dual $V^\mu\equiv\tfrac16\epsilon^{\mu\nu\rho\sigma}H_{\nu\rho\sigma}=A_\nu\Ft^{\mu\nu}+\tfrac12\epsilon^{\mu\nu\rho\sigma}\partial_\nu B_{\rho\sigma}$
is precisely the combination coupled to $j_\mu$, and obeys $\partial_\mu V^\mu=\tfrac12F\Ft$.
In terms of $V^\mu$,
\begin{equation}
\mathcal L=-\frac14F^2-\frac{1}{2g_B^2}V_\mu V^\mu+\kappa\,j_\mu V^\mu,
\qquad |\kappa|=\frac{b_J}{2M^2}=\frac1{2\Lstar^2},
\label{eq:KR}
\end{equation}
where $g_B$ normalizes the 2-form kinetic term. Note that $V_\mu V^\mu$ plays the role of the kinetic term of the 2-form so the limit $g_B\to\infty$ in Ref.~\cite{Geng:2007va} corresponds to a non-dynamical 2-form.

\textit{Duality.}---Locally every divergence-free vector is of the form $\tfrac12\epsilon^{\mu\nu\rho\sigma}\partial_\nu B_{\rho\sigma}$, so integrating over $B$ is equivalent to integrating over $V$ subject to $\partial_\mu V^\mu=\tfrac12F\Ft$. Imposing the constraint with a multiplier $a$ and integrating by parts,
\begin{equation}
\mathcal L=-\frac14F^2-\frac{1}{2g_B^2}V_\mu V^\mu+(\kappa j_\mu-\partial_\mu a)V^\mu-\frac a2F\Ft .
\end{equation}
$V_\mu$ is now auxiliary. Eliminating it, $V_\mu=g_B^2(\kappa j_\mu-\partial_\mu a)$, and setting $\varphi=g_B\,a$,
\begin{equation}
\mathcal L=-\frac14F^2+\frac12\partial\varphi^2-\frac{g_{a\gamma}}4\varphi F\Ft
+\frac{\partial_\mu\varphi\,j^\mu}{f_\ell}+\frac{j_\mu j^\mu}{2f_\ell^2},
\label{eq:dual}
\end{equation}
with $g_{a\gamma}=2/g_B$ and $1/f_\ell=g_B|\kappa|$. The gauge invariance $B\to B+d\Lambda$, where $\Lambda$ is a 1-form, forbids any potential for $B$, so $\varphi$ is massless (up to non-perturbative effects). This is the Chern--Simons (or axion) branch of 2-form/axion duality, familiar from the model-independent axion of string theory~\cite{Svrcek:2006yi}. Note that the normalization $g_B$ cancels in the ratio
\begin{equation}
\frac{g_{a\gamma}}{f_\ell}=\frac{b_J}{M^2}=\frac{1}{\Lstar^2},
\label{eq:product}
\end{equation}
which is exactly the combination that appears in the birefringence angle. 

The dual theory reproduces the signal: on the background, with $R$ the scale factor and dots denoting derivatives with respect to $t$, $\tfrac{d}{dt}[R^3(\dot a-\kappa\Delta n_{\nu_e})]=0$, so for vanishing Kalb--Ramond flux, $V^0=0$, we have $\dot a=\kappa\Delta n_{\nu_e}$. Using $-\tfrac a2F\Ft=-a\,\partial_\mu K^\mu$ and integrating by parts, we obtain  Eq.~\eqref{eq:LJ}. The same result can be obtained using the usual form of the birefringence angle for an axion coupled to photons: $\beta = \frac{1}{2}g_{a\gamma}g_B\int \dot{a}dt = |\kappa|\int \Delta n dt$, reproducing Eq.\eqref{eq:bire_angle}. The neutrino-current model is an axion-rotation model~\cite{Co:2026rot} in disguise (see also \cite{Minotti:2026bkc}).

\textit{The photon mass.}---Since the 2-form field $B$ shifts under $U(1)_{\rm em}$, one might
expect its dual, $a$, to shift and to be eaten by the photon. However, it is not the case. $V_\mu$ and $j_\mu$ are gauge invariant under EM so $\partial_\mu a$ is gauge invariant, too. Duality exchanges equations of motion and Bianchi identities, so the shift of $B$ becomes the coupling to photons, $aF\Ft$. A coupling $B\wedge F$ would have produced a shift of $a$ under EM and, hence, a St\"uckelberg mass for the photon~\cite{Cremmer:1973mg} (see also \cite{Platschorre:2024xxp}). That term is forbidden here: under a EM transformation, $\delta B=-\chi F$, and hence $\delta\int B\wedge F=-\int\chi F\wedge F$, which is not a total derivative. If that term was present, only an electromagnetic anomaly could cancel it, as in the Green--Schwarz mechanism~\cite{Green:1984sg}. The anomaly requires fermions chiral under $U(1)_{\rm em}$, which are excluded. The same argument forbids a mass for $B$ through a coupling to a new dark U(1) sector~\cite{Platschorre:2024xxp}, $B\wedge F'$, which would imply that $\varphi$ is eaten. As before, $\delta\int B\wedge F'=-\int\chi\,F\wedge F'$ is not a total derivative and leads to an inconsistent theory unless new fermions carry a mixed $U(1)'\times U(1)_{\rm em}^2$ anomaly, in which case $\varphi$ is eaten by a massive gauge boson $A'$. (The $B\wedge F$ term discussed in Ref.~\cite{Nakai:2023zdr} applies to a 2-form inert under $U(1)_{\rm em}$.) The only new light state is therefore the axion $\varphi$.

\textit{Electron coupling.}---$\varphi$ is an SM singlet, so $j^\mu$ must be an $SU(2)_L\times U(1)_Y$ singlet. Without Higgs insertions, the singlet currents containing $\bar\nu_e\gamma^\mu\nu_e$ are
\begin{equation}
j^\mu=\bar L_e\gamma^\mu L_e+c_R\,\bar e_R\gamma^\mu e_R\supset-\frac{1-c_R}{2}\,\bar e\gamma^\mu\gamma_5e ,
\end{equation}
where $L_e=(\nu_{eL},e_L)$ is the electron lepton doublet, $e_R$ the right-handed electron and $c_R$ a real coefficient, with $c_R=0$ for the left-handed current of Refs.~\cite{Geng:2007va,Zhou:2023aqh,Yin:2026nu}. Classically $\partial_\mu j^\mu=-(1-c_R)\,m_e\,\bar e\,i\gamma_5e$, so only the axial electron term contributes to $\partial_\mu\varphi\,j^\mu$ at tree level. This gives the usual axion-electron coupling $\mathcal L\supset(g_{ae}/2m_e)\partial_\mu\varphi\,\bar e\gamma^\mu\gamma_5e$ with $g_{ae}=(1-c_R)m_e/f_\ell$, and Eq.~\eqref{eq:product} fixes the product
\begin{equation}
g_{a\gamma}\,g_{ae}=(1-c_R)\frac{m_e}{\Lstar^2}
=(1-c_R)\,\frac{2.6\times10^{-13}}{\xi_{\nu_e}}\GeV^{-1},
\label{eq:gagge}
\end{equation}
i.e.\ $5.3\times10^{-12}\GeV^{-1}$ for $\xi_{\nu_e}=0.05$ and $c_R=0$.

\begin{figure}[t]
\centering
\includegraphics[width=\columnwidth]{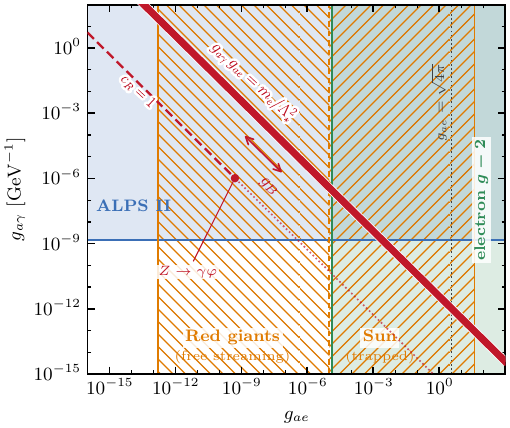}
\caption{The prediction Eq.~\eqref{eq:gagge} for $c_R=0$ and $0.05\le\xi_{\nu_e}\le0.1$ (red band) in the $(g_{ae},g_{a\gamma})$ plane; the unobserved normalization $g_B$ moves the model along the band (arrow). Shaded regions are excluded in the laboratory, by ALPS~II~\cite{ALPSII:2025eri} and by the electron $g-2$ ($|\Delta a_e|\lesssim10^{-12}$); hatched regions are excluded by stars, by red giants~\cite{Capozzi:2020cbu} where axions stream freely and by the Sun where they are trapped, $\lambda_\gamma\ll\lambda_a\ll R_\odot$ (see Supplemental Material). The dashed line separates the two regimes and the dotted line marks $g_{ae}=\sqrt{4\pi}$. No point of the band survives. The dashed red line corresponds to  $c_R=1$ for $\xi_{\nu_e}=0.05$, with only the loop-induced $g_{ae}\sim10^{-4}m_e/f_\ell$; below the dot, where it turns dotted, $Z\to\gamma\varphi$ excludes it (see Supplemental Material) and above it ALPS~II does it, too.}
\label{fig:exclusion}
\end{figure}

\textit{Exclusion.}---As shown in Fig.~\ref{fig:exclusion}, laboratory data alone exclude the couplings given in Eq.~\eqref{eq:gagge}.  ALPS~II experiment bounds the axion-photon coupling $g_{a\gamma}<1.5\times10^{-9}\GeV^{-1}$ for a light axion~\cite{ALPSII:2025eri}, so Eq.~\eqref{eq:gagge} requires $g_{ae}\gtrsim1.8\times10^{-3}$ for $\xi_{\nu_e}=0.1$, near the largest asymmetry allowed by big-bang nucleosynthesis (BBN)~\cite{Steigman:2012ve}. A massless or light pseudoscalar shifts the electron anomalous magnetic moment $a_e=(g_e-2)/2$ by $\Delta a_e=-g_{ae}^2/16\pi^2$~\cite{Leveille:1977rc,Lindner:2016bgg}, i.e.\ by $-2.0\times10^{-8}$ in the case of $g_{ae}\sim1.8\times10^{-3}$. The measurement~\cite{Fan:2022eto}, however, agrees with the SM prediction at the level of $\sim10^{-12}$ for the two determinations of the fine-structure constant~\cite{Parker:2018vye,Morel:2020dww}. Therefore, this measurement rules out the axion-electron coupling that the signal requires.

Stars provide an independent and even stronger constraint. Eq.~\eqref{eq:gagge} exceeds the helioscope bound $g_{a\gamma}g_{ae}<8.1\times10^{-23}\GeV^{-1}$~\cite{Barth:2013sma} by eleven orders of magnitude, and the two couplings cannot both satisfy $g_{a\gamma}<5.8\times10^{-11}\GeV^{-1}$~\cite{CAST:2024eil} and $g_{ae}<1.6\times10^{-13}$~\cite{Capozzi:2020cbu}. These bounds, however, assume that axions leave efficiently the stars. For relatively large couplings, axions may be \textit{trapped} but this does not rescue the model (see supplemental material S1 for more details). Similar to photons, axions may be absorbed due to their electron coupling. The relevant cross sections are rescaled by $\simeq$$\,3g_{ae}^2\omega^2/16\pi\alpha m_e^2$~\cite{Pospelov:2008jk}, where $\omega$ is the energy and $\alpha$ the fine-structure constant. In the solar core this gives a ratio of the axion and photon mean free paths $\lambda_a/\lambda_\gamma\simeq9\times10^{10}(1.5\times10^{-4}/g_{ae})^2$, while $\lambda_a$ is smaller than the solar radius $R_\odot$ for $g_{ae}\gtrsim10^{-5}$. Axions that are trapped still interact more weakly than photons and raise the  conductivity by a factor $\sim\lambda_a/\lambda_\gamma$. They escape from inside the star with an axion luminosity $L_a$ far above the solar luminosity, $L_a/L_\odot\gg1$. Trapping protects a star only if $\lambda_a\lesssim\lambda_\gamma$~\cite{Raffelt:1996wa}, which in this case requires $g_{ae}\gg 1$. Stars thus exclude $10^{-13}\lesssim g_{ae}\lesssim \mathcal{O}(1)$, the lower end from red giants and the upper one from the Sun, and with it the whole perturbative range of $g_{ae}$ allowed by Eq.~\eqref{eq:gagge}. Independently of stellar physics, the trapping regime is also excluded by the electron $g-2$: $|\Delta a_e|=g_{ae}^2/16\pi^2\lesssim10^{-12}$ requires $g_{ae}\lesssim1.3\times10^{-5}$ for any $g_{a\gamma}$, which is where the trapping regime becomes relevant.

\textit{Other currents.}---The ALPS II bound and Eq.~\eqref{eq:product} together imply
\begin{equation}
f_\ell=g_{a\gamma}\Lstar^2\lesssim0.3\GeV .
\label{eq:fl}
\end{equation}
Similar bounds apply to other flavors. For example, in the muon case and for $c_R=0$, an interaction with the $\nu_\mu$ current gives $g_{a\mu}=m_\mu/f_\ell\gtrsim0.35$ and hence, similar to the electron case we have $|\Delta a_\mu|=g_{a\mu}^2/16\pi^2\gtrsim8\times10^{-4}$. The measurement of the muon anomalous magnetic moment~\cite{Muong-2:2025xyk} agrees with the SM prediction~\cite{Aliberti:2025beg}, $a_\mu^{\rm exp}-a_\mu^{\rm SM}=(38\pm63)\times10^{-11}$. Therefore, the measurement excludes this shift by six orders of magnitude. 

There are in principle two ways to remove the tree-level electron coupling. In this section we show that even if the tree-level coupling is removed, Eq.~\eqref{eq:fl} still excludes the model. For $c_R=1$, the current $j_\mu$ is the electron lepton number current. It is anomalous under the weak $SU(2)_L$, so $\partial_\mu\varphi\,j^\mu/f_\ell=-\varphi\,\partial_\mu j^\mu/f_\ell$ contains the couplings to gauge bosons
\begin{equation}
    -\frac{\varphi}{32\pi^2f_\ell}\,(g_2^2W^a\widetilde W^a-g_1^2Y\widetilde Y)\,.
\end{equation}
Here $g_2$, $g_1$ are the $SU(2)_L$ and $U(1)_Y$ gauge couplings and $W^a_{\mu\nu}$, $Y_{\mu\nu}$ the corresponding field strengths. The coupling to photons cancels, but the coupling $-(\alpha/4\pi s_wc_wf_\ell)\,\varphi\,Z_{\mu\nu}\Ft^{\mu\nu}$ survives, with $s_w$, $c_w$ the sine and cosine of the weak mixing angle, and $Z_{\mu\nu}$ the $Z$ field strength. This interaction gives a decay rate $\Gamma(Z\to\gamma\varphi)\simeq22\,\mathrm{MeV}\,(\GeV/f_\ell)^2$ (see Supplemental Material), and branching ratio $\text{BR}\gtrsim0.1$ when using Eq.~\eqref{eq:fl}. L3 bounds single-photon $Z$ decays into invisible particles at the $10^{-6}$ level~\cite{L3:1997exg}, which here requires $f_\ell\gtrsim100\GeV$. Electroweak boson loops also generate a coupling to the electron $g_{ae}\sim10^{-4}m_e/f_\ell\gtrsim10^{-7}$~\cite{Bauer:2020jbp}, which lies inside the range excluded by stars. 

One could instead dress the current with the Higgs field so that it contains only neutrinos. Such a current is not conserved, because the $W$ boson turns $\nu_e$ into $e$, so $-\varphi\,\partial_\mu j^\mu/f_\ell$ contains a contact coupling $\frac{\varphi}{f_\ell}\,W^+_\mu\bar\nu_L\gamma^\mu e_L$. With the value in Eq.~\eqref{eq:fl}, the decays of the $W$ and Higgs bosons into $\varphi$ exceed the SM rates implying that models with such modified currents are also excluded.

\textit{Conserved currents.}---The arguments in the previous section rely on $\partial_\mu j^\mu\neq0$. A current that is anomaly-free, such as $j=\mathrm{B}-\mathrm{L}$, evades them (see Supplemental Material). In this case we still have $j^0\simeq\Delta n_\nu$, but $\partial_\mu\varphi\,j^\mu$ is now a total derivative, so $\varphi$ is an axion coupled only to photons (see Eq.~\eqref{eq:dual}). The neutrino asymmetry then fixes the axion field velocity when the asymmetry is created. The Bianchi identity conserves $R^3V^0=g_B^2R^3(\kappa j^0-\dot a)$. For vanishing flux, $V^0=0$, we have $\dot a=\kappa j^0$, and the axion evolves freely afterwards. The scenario is then an axion-rotation model where the birefringence angle is induced via the photon coupling~\cite{Co:2026rot} (see also~\cite{Minotti:2026bkc}), with the cosmic neutrinos present at recombination playing
no role. The only remaining trace of the current is the contact term in
Eq.~\eqref{eq:dual}, $\frac{j_\mu j^\mu}{2f_\ell^2}$. As for the other currents, the ALPS~II bound gives $f_\ell\lesssim0.3\GeV$, Eq.~\eqref{eq:fl}. The contact term is then a four-fermion interaction with strength $\gtrsim5\times10^5G_F$ (where $G_F$ is the Fermi constant) among electrons,
quarks and neutrinos. Such interactions are excluded by neutrino--electron and neutrino--quark scattering~\cite{Farzan:2017xzy}
as well as LEP contact-interaction searches~\cite{ALEPH:2013dgf}. 
Table~\ref{tab:currents} summarizes the currents considered and the processes/experiments that exclude them.

\begin{table}[t]
\renewcommand{\arraystretch}{1.2}\small
\begin{tabular}{@{}lll@{}}
\toprule
current $j$ & $\partial_\mu j^\mu$ & excluded by\\
\midrule
$\bar L_e\gamma^\mu L_e$ ($c_R=0$) & $-m_e\bar e i\gamma_5e$ & $g-2$, ALPS~II, stars\\
$\mathrm{L}_e$ ($c_R=1$) & EW anomaly & $Z\to\gamma\varphi$\\
Higgs-dressed & $\propto W^+_\mu\bar\nu\gamma^\mu e_L$ & $W$, $h$ decays\\
$\mathrm{B}-\mathrm{L}$ & $0$ & contact term\\
\bottomrule
\end{tabular}%
\caption{Different currents $j_\mu$ coupled to the KR field, the divergence that sets the couplings of $\varphi$ to SM fields through $-\varphi\,\partial_\mu j^\mu/f_\ell$, and what excludes each case for $f_\ell\lesssim0.3\GeV$.}
\label{tab:currents}

\end{table}

\textit{The alternative.}---Gauge invariance could also be restored via the St\"uckelberg mechanism with the photon shifting as
\begin{equation}
    A_\mu\to A_\mu-\partial_\mu\pi/m_\gamma\,,
\end{equation}
with $\pi$ a Goldstone boson eaten by the photon. In this case the photon becomes massive, and $\pi$ couples as
\begin{equation}
   \frac{1}{2m_\gamma\Lstar^2} \pi\,(\partial_\nu j_\mu)\Ft^{\mu\nu}\,.
\end{equation}
This interaction becomes strongly coupled at energy scales $E\sim(m_\gamma\Lstar^2)^{1/3}$. This resembles the model presented in~\cite{Kamada:2026tfd} where a 3-form is coupled to photons. The difference is that in that case, the birefringence signal fixes the
dimensionless coupling between the 3-form and photons to be of order $H_0/M_p$, with $M_p$ the Planck mass. 
Here, instead, the UV scale that sets the coupling strength (see Eq.~\eqref{eq:product}) is fixed at $\Lstar\simeq10$\,TeV. If one uses the bounds to the photons mass from Coulomb's law~\cite{Williams:1971ms}, $m_\gamma\lesssim10^{-14}$\,eV, and from the solar wind~\cite{Ryutov:2007zz,ParticleDataGroup:2024cfk}, $10^{-18}$\,eV, then we find that the effective theory would break down at energies larger than $\simeq \mathcal{O}(1)$\,keV. This implies that the EFT is not valid at BBN temperatures, where the asymmetry is set. For this reason, we will not explore this possibility further.

\textit{Conclusion.}---Cosmic neutrinos cannot induce a cosmic birefringence signal.
The reason is that the gauge-invariant UV completion used in the literature -- a Kalb--Ramond 2-form coupled as in Eq.~\eqref{eq:KR} -- is dual to a massless axion coupled to photons and leptons. The product of its couplings to photons and to the lepton current is fixed by the signal, as shown in Eq.~\eqref{eq:product}. $SU(2)_L$ invariance then extends the axion coupling to electrons, and the required value for the product $g_{a\gamma}g_{ae}$ is excluded by laboratory data and stars, see Eq.~\eqref{eq:gagge} and Fig.~\ref{fig:exclusion}. Models with currents without a tree-level electron coupling (i.e. it cancels due to axion charge assignments) behave in a similar way. Anomalous or Higgs-dressed currents lead to $Z$, $W$ or Higgs decays into axions far above current limits. For exactly conserved currents, either a contact interaction excludes the model or the relic neutrinos decouple from the signal. A St\"uckelberg completion instead gives the photon an excessively large mass and/or is not under control at BBN, excluding this as a consistent option. A cosmic birefringence signal should therefore be read as evidence for a light axion coupled to photons and not as a probe of the relic-neutrino asymmetry.

The discovery of a birefringent axion would be much more than discovering a new particle: it would offer a unique window on the ultraviolet limit of the Standard Model, far beyond the reach of other experiments. With $g_{a\gamma}=\mathcal C\alpha/2\pi f$ and $\Delta\theta=\Delta\varphi/f$, the rotation angle $\beta=\mathcal C\,\alpha\,\Delta\theta/4\pi$ is independent of $f$ and requires $\mathcal C\,\Delta\theta\simeq \mathcal{O}(1-10)$. The required axion mass, $m\lesssim10^{-28}$\,eV to generate an isotropic signal, requires a shift symmetry of extraordinary quality. Quantum gravity, however, is expected to admit no exact global symmetries~\cite{Banks:2010zn,Harlow:2018jwu}. If instantons of action $S$ generate the potential, naively of the form $V(\theta)\sim m^2f^2\cos\theta\sim e^{-S}M_{\rm UV}^4\cos\theta$, the signal requires $S\gtrsim250$ (here $M_{\rm UV}$ is a UV scale typically related to the UV cutoff of the EFT). As the axion weak gravity conjecture relates the instanton action and the decay constant, $fS\lesssim M_P$~\cite{Arkani-Hamed:2006emk,Svrcek:2006yi}, then we obtain $f\lesssim10^{16}$\,GeV, so that birefringence is in some sense a test of quantum gravity principles~\cite{Alvey:2020nyh,Alvey:2021hjp,DAmico:2026fuy}. Finally, independently of the size of $f$, gauge unification ties the photon coupling of an axion to its gluon coupling~\cite{Agrawal:2022lsp,Agrawal:2025rbr,Agrawal:2024ejr,Reig:2025dqb,Nee:2026msa}. A birefringent axion would provide a robust test whether, and how, the SM is unified in the far UV.

\begin{acknowledgments}
Acknowledgments: I thank Andrea Caputo, Miguel Escudero, Silvia Gasparotto, Eiichiro Komatsu, Arthur Platschorre, Ryan Plestid and Vivian Poulin for useful discussions and comments on the manuscript. 
Claude Opus 5.5 (Anthropic) was used for language editing, plotting Fig.~\ref{fig:exclusion}, and adversarially auditing the results in the manuscript. The author checked all AI output, including every equation and numerical estimate, and takes full responsibility for the content.

\end{acknowledgments}

\clearpage
\setcounter{equation}{0}
\renewcommand{\theequation}{S\arabic{equation}}
\setcounter{secnumdepth}{1}
\setcounter{section}{0}
\renewcommand{\thesection}{S\arabic{section}}
\begin{center}
\textbf{\large Supplemental Material}
\end{center}

\section{Stellar evolution: free streaming versus trapping}
\label{sec:stellar}

Helioscope and stellar bounds on axions assume that axions produced in a star leave it freely. For the couplings required here this is not necessarily the case. In this section we consider cases where axions do not leave the star efficiently. The conclusion is that trapping does not help: the Sun excludes the model for any perturbative $g_{ae}$. All the estimates are approximate, using the solar centre of the standard solar model~\cite{Bahcall:2004pz}: $T_c\simeq1.35$\,keV, $\rho_c\simeq150\,\mathrm{g\,cm^{-3}}$
for the central temperature and density, and a Rosseland opacity
$\mathcal K_\gamma\sim1\,\mathrm{cm^2g^{-1}}$, so that the photon mean free path is
$\lambda_\gamma=1/(\mathcal K_\gamma\rho)\sim10^{-2}$\,cm, with $\rho$ the local density. We write $\mathcal K_a$
and $\lambda_a=1/(\mathcal K_a\rho)$ for the corresponding axion quantities.

\paragraph{Axion opacity.} Due to their coupling to electrons, $g_{ae}$, axions are absorbed by the same electron processes that generate the photon opacity. Up to factors of order unity, the axion and photon absorption cross sections, $\sigma_a$ and $\sigma_\gamma$, are related by~\cite{Pospelov:2008jk}
\begin{equation}
\frac{\sigma_a}{\sigma_\gamma}\simeq\frac{3\,g_{ae}^2\,\omega^2}{16\pi\alpha\,m_e^2},
\end{equation}
with $\omega$ the energy, $\alpha$ the fine-structure constant and $m_e$ the electron mass. Taking a typical thermal energy $\omega\simeq3T_c\simeq4$\,keV,
\begin{equation}
\frac{\lambda_a}{\lambda_\gamma}=\frac{\mathcal K_\gamma}{\mathcal K_a}
\simeq9\times10^{10}\Big(\frac{1.5\times10^{-4}}{g_{ae}}\Big)^2\Big(\frac{4\,\mathrm{keV}}{\omega}\Big)^2 .
\label{eq:lamratio}
\end{equation}

\paragraph{The three regimes.} Comparing $\lambda_a$ with the solar radius $R_\odot$ and with $\lambda_\gamma$:
\begin{enumerate}
\item \emph{Free streaming}, $\lambda_a\gtrsim R_\odot$, i.e.\ $g_{ae}\lesssim10^{-5}$. Axions
escape and carry away energy at a rate $\propto g_{ae}^2$. This is the regime of the
standard bounds, $g_{ae}\lesssim10^{-13}$ from red giants~\cite{Capozzi:2020cbu}; the
Sun itself requires $g_{ae}$ many orders of magnitude below $10^{-5}$.
\item \emph{Weak trapping with} $\lambda_\gamma\ll\lambda_a\ll R_\odot$, i.e.\
$10^{-5}\lesssim g_{ae}\lesssim40$. Axions are in local thermal equilibrium, but they
interact much more weakly than photons.
\item \emph{Strong trapping}, $\lambda_a\lesssim\lambda_\gamma$, i.e.\ $g_{ae}\gtrsim40$, far
outside the perturbative range. 
\end{enumerate}
The model lives in regime 2: with $g_{a\gamma}g_{ae}=m_e/\Lambda_*^2$ (Eq.~\eqref{eq:gagge} of the Letter), the laboratory bound $g_{a\gamma}<1.5\times10^{-9}\GeV^{-1}$ requires $g_{ae}\gtrsim2\times10^{-3}$, so that throughout the perturbative range $\lambda_a/\lambda_\gamma\sim 10^2$--$10^9$.

\paragraph{Why trapping does not help.} In the regime 2, energy is carried by diffusion of both photons and axions. A massless boson with one polarization has half the energy density of photons, so at a given temperature gradient the total energy flux $\mathcal F$ is related to the radiative flux $\mathcal F_\gamma$ by
\begin{equation}
\mathcal F=\mathcal F_\gamma\Big(1+\frac{\lambda_a}{2\lambda_\gamma}\Big),
\end{equation}
i.e.\ the conductivity of the stellar interior grows by $\lambda_a/2\lambda_\gamma$. The axions eventually escape from an ``axiosphere'' where their optical depth is $\tau_a\simeq2/3$. Since $\tau_a\simeq\tau_\gamma\,\lambda_\gamma/\lambda_a$, this happens where the photon optical depth is $\tau_\gamma\simeq\tfrac23\lambda_a/\lambda_\gamma\gg1$, i.e.\ deep inside the star and at a temperature much higher than the photospheric one. Using the grey-atmosphere relation $T^4\simeq\tfrac34T_{\rm eff}^4(\tau_\gamma+\tfrac23)$, with $T_{\rm eff}$ the effective
surface temperature, as a rough guide, the axion luminosity $L_a$ from an axiosphere of radius $r_a$, in units of the solar luminosity $L_\odot$, is
\begin{equation}
\frac{L_a}{L_\odot}\sim\frac14\,\frac{\lambda_a}{\lambda_\gamma}\Big(\frac{r_a}{R_\odot}\Big)^2 ,
\end{equation}
which for $\lambda_a/\lambda_\gamma\sim 10^2$--$10^9$ exceeds the solar luminosity by a factor $\sim40$ at $g_{ae}=\sqrt{4\pi}$ and by many orders of magnitude below. Whether one phrases it as a vastly enhanced energy loss or as a vastly enhanced conductivity, the Sun could not exist in its observed state.

\paragraph{The criterion.} Trapping protects a star when axions carry less energy than radiation, which here occurs when they interact \emph{more} strongly than photons and $\lambda_a\lesssim\lambda_\gamma$~\cite{Raffelt:1996wa}. In our case that would require $g_{ae}\gtrsim40$, which corresponds to $f_\ell$ a factor $\sim 40$ smaller than the electron mass. 
Regime 1 and 2 exclude the whole range $10^{-13}\lesssim g_{ae}\lesssim40$, and therefore the model: by Eq.~\eqref{eq:gagge} of the Letter, $g_{ae}<10^{-13}$ would require $g_{a\gamma}\gtrsim 20\GeV^{-1}$, which laboratory searches exclude by many orders of magnitude. The standard free-streaming bounds thus do not apply directly, but stellar evolution still provides the strongest constraints. Furthermore, the trapping regimes 2 and 3 are excluded by laboratory experiments alone: the electron $g-2$ bound $|\Delta a_e|=g_{ae}^2/16\pi^2\lesssim10^{-12}$ requires $g_{ae}\lesssim1.3\times10^{-5}$, independently of $g_{a\gamma}$.

\section{The case $c_R=1$}
\label{sec:cR1}

For $c_R=1$ the current $j=\bar L_e\gamma^\mu L_e+\bar e_R\gamma^\mu e_R$ is the electron lepton number, so the tree-level axion--electron coupling vanishes, see Eq.~\eqref{eq:gagge}. The current is, however, anomalous with respect to the electroweak sector. With $\widetilde X^{\mu\nu}=\tfrac12\epsilon^{\mu\nu\rho\sigma}X_{\rho\sigma}$ and for general $c_R$,
\begin{equation}
\partial_\mu j^\mu\supset\frac{1}{32\pi^2}\Big[g_2^2\,W^a\widetilde W^a+(1-2c_R)\,g_1^2\,Y\widetilde Y\Big],
\end{equation}
where $g_2$, $g_1$ are the $SU(2)_L$ and $U(1)_Y$ gauge couplings, $W^a_{\mu\nu}$, $Y_{\mu\nu}$ the corresponding field strengths, and $X$ stands for any of them. With $s_w$, $c_w$ the sine and cosine of the weak mixing angle and $e=g_2s_w=g_1c_w$, writing $W^3=c_wZ+s_wA$ and $Y=-s_wZ+c_wA$, the
neutral part is
\begin{align}
&\frac{e^2}{32\pi^2}\Big[2(1-c_R)\,F\tilde F+\frac{2\,(c_w^2-(1-2c_R)s_w^2)}{s_wc_w}\,Z\tilde F
\\&+\frac{c_w^4+(1-2c_R)s_w^4}{s_w^2c_w^2}\,Z\tilde Z\Big].
\end{align}
The photon term cancels for $c_R=1$, consistently with the current being vector-like with respect to EM, but the $Z\tilde F$ term does not. Through $\partial_\mu\varphi\,j^\mu/f_\ell=-\varphi\,\partial_\mu j^\mu/f_\ell$ one obtains, for $c_R=1$,
\begin{equation}
\mathcal L\supset-\frac{\alpha}{4\pi s_wc_w\,f_\ell}\,\varphi\,Z_{\mu\nu}\tilde F^{\mu\nu} .
\end{equation}
For $\mathcal L=-C\,\varphi Z_{\mu\nu}\tilde F^{\mu\nu}$, with $C$ a constant, and a massless $\varphi$, the rate is $\Gamma(Z\to\gamma\varphi)=C^2m_Z^3/24\pi$, so with $\alpha(m_Z)=1/128$ we obtain
\begin{align}
&\Gamma(Z\to\gamma\varphi)\simeq22\,\mathrm{MeV}\left(\frac{1\GeV}{f_\ell}\right)^2,\\&
\mathrm{BR}\simeq9\times10^{-3}\left(\frac{1\GeV}{f_\ell}\right)^2 .
\end{align}
For $f_\ell\lesssim0.3\GeV$ (see Eq.~\eqref{eq:fl}) this gives $\mathrm{BR}\gtrsim0.1$. The decay produces a monochromatic photon with $E_\gamma=m_Z/2$ and missing energy (note that the axion $\varphi$ is not expected to interact with the detector despite $f_\ell$ is small). L3 bounds such single-photon $Z$ decays to $\mathrm{BR}(Z\to\gamma X)\lesssim10^{-6}$ for $E_\gamma\approx m_Z/2$~\cite{L3:1997exg}, which requires $f_\ell\gtrsim100\GeV$. $Z$ decays therefore exclude the case $c_R=1$.

Finally, we also note that electroweak boson loops regenerate an electron coupling at two loops~\cite{Bauer:2020jbp},
$g_{ae}\sim(\alpha_2/4\pi)^2\ln(\Lambda_*^2/m_W^2)\,m_e/f_\ell\sim10^{-4}\,m_e/f_\ell$, with $\alpha_2=\alpha/s_w^2$. It assumes $g_{ae}=0$ at $\Lambda_*$ motivated by the choice $c_R=1$. The estimate then provides a minimal axion-electron coupling at the level of $g_{ae}\gtrsim10^{-7}$. This value lies
inside the window excluded by stars, see Fig.~\ref{fig:exclusion} for details.

\section{Conserved, anomaly-free currents}
\label{sec:conserved}

The couplings to electrons discussed in the sections above come from $\partial_\mu j^\mu\neq0$: the axial electron current at tree level, the $SU(2)_L$ anomaly for $c_R=1$, and the charged current for Higgs-dressed currents. A singlet current containing $\bar\nu_e\gamma^\mu\nu_e$ that is vector-like, exactly conserved and anomaly-free might evade all three. In this section we provide the details about how this option is excluded due to a contact 4-fermion interaction between SM fermions.

\paragraph{Example.} A current of this type is
\begin{equation}
j=\mathrm{B}-\mathrm{L} ,
\label{eq:conserved}
\end{equation}
where $\mathrm{L}$ is total lepton number and $\mathrm{B}$ is baryon number. It is conserved in the SM with massless neutrinos, and its gauge anomalies cancel. 
Majorana neutrino masses also break $\mathrm{B}-\mathrm{L}$; the resulting couplings are proportional to $m_\nu/f_\ell$ and involve neutrinos only.

On the cosmological background, with $n_e$, $n_p$, $n_n$, $n_{\rm L}$ and $n_{\rm B}$ the electron, proton,
neutron, lepton-number and baryon-number densities, charge neutrality ($n_e=n_p$) gives
\begin{equation}
j^0=n_{\rm B}-n_{\rm L}=-\Big(\sum_\alpha\Delta n_{\nu_\alpha}-n_n\Big) ,
\end{equation}
and for $\xi_\nu\sim0.05$ the neutron term is $\sim10^{-8}$ of the first. The signal is therefore
unchanged, up to a sign. 

\paragraph{The axion decouples from the current.} With $\partial_\mu j^\mu=0$, the coupling $\partial_\mu\varphi\,j^\mu/f_\ell$ is a total derivative, and the dual Lagrangian reduces to
\begin{equation}
\mathcal L=-\frac14F^2+\frac12(\partial\varphi)^2-\frac{g_{a\gamma}}4\varphi F\tilde F+\frac{j_\mu j^\mu}{2f_\ell^2} .
\label{eq:dualcons}
\end{equation}
Because $j_\mu$ is exactly conserved (up to small neutrino masses), this holds to all orders: no loop regenerates a coupling of $\varphi$ to electrons, $W$ or $Z$ bosons. $\varphi$ is a free axion coupled only to photons.

\paragraph{Origin of the birefringence signal.} In the 2-form language, the modified Bianchi identity $\partial_\mu V^\mu=\tfrac12F\tilde F$ gives on the background $\tfrac{d}{dt}(R^3V^0)=0$, with $R$ the scale factor and $V^0=g_B^2(\kappa j^0-\dot a)$. Hence $R^3(\kappa j^0-\dot a)$ is constant for any current, independently of whether it is conserved or not. For a conserved current $R^3j^0$ is constant as well, so $R^3\dot a$ is constant. The axion evolves freely, and its velocity redshifts as $R^{-3}$, just like the asymmetry. A nonzero $R^3j^0$ can only be created by $j$-violating dynamics in the early universe, such as heavy Majorana neutrinos or an Affleck--Dine condensate. If the flux vanishes, $V^0=0$, at that time, these dynamics give the axion exactly $\dot a=\kappa j^0$. After that, the axion and the neutrinos are completely decoupled. The rotation observed in the CMB is that of a free massless axion whose velocity was set at the time the asymmetry formed, i.e.\ an axion-rotation model~\cite{Co:2026rot}. The cosmic neutrinos present at recombination do not
participate in generating the birefringence.

\paragraph{The constraint: the photon coupling and the contact term.} Similar to the other currents above, we use laboratory constraints. With
$f_\ell=g_{a\gamma}\Lambda_*^2$, the ALPS~II bound $g_{a\gamma}<1.5\times10^{-9}\GeV^{-1}$~\cite{ALPSII:2025eri}
gives $f_\ell\lesssim0.3\GeV$ for $\xi_{\nu_e}\le0.1$ ($0.15\GeV$ for $\xi_{\nu_e}=0.05$), Eq.~\eqref{eq:fl}. Since $g_{ae}=0$,
nothing traps the axion in stars, so the free-streaming bounds of CAST~\cite{CAST:2024eil} and SN1987A~\cite{Payez:2014xsa}  apply. They provide an independent check and would strengthen this to $f_\ell\lesssim11$\,MeV and $1$\,MeV for $\xi_{\nu_e}\le0.1$.

Using the ALPS~II bound, the contact term in Eq.~\eqref{eq:dualcons} then has strength
\begin{equation}
\frac{1}{2f_\ell^2}\gtrsim5.6\GeV^{-2}\simeq5\times10^5\,G_F
\label{eq:contact_coeff}
\end{equation}
for $\xi_{\nu_e}\le0.1$, with $G_F$ the Fermi constant. This arises from tree-level exchange of the 2-form. For a conserved current the exchange is exactly local, $\tfrac12g_B^2\kappa^2\,j^\mu(\eta_{\mu\nu}-k_\mu k_\nu/k^2)j^\nu\to j^2/2f_\ell^2$, with $k$ the momentum transfer and $\eta_{\mu\nu}$ the Minkowski metric,
so it applies at all momenta.

For $j=\mathrm{B}-\mathrm{L}$ this contact interaction contains $(\bar e\gamma e)^2$, $(\bar e\gamma e)(\bar q\gamma q)$ and $(\bar\nu\gamma\nu)(\bar e\gamma e)$.
LEP contact-interaction searches require such 4-fermion operators to have coefficients $\lesssim4\pi/(10\,\mathrm{TeV})^2\simeq10^{-7}\GeV^{-2}$~\cite{ALEPH:2013dgf}, excluding Eq.~\eqref{eq:contact_coeff} by several orders of magnitude. Neutrinos interacting with
electrons $5\times10^5$ times more strongly than through the weak force would also stay in
equilibrium with the plasma until $e^+e^-$ annihilation is complete ($T\sim20$\,keV), sharing its entropy. This gives
$N_{\rm eff}\simeq3\,(11/4)^{4/3}\approx12$, in conflict with BBN and with the value $N_{\rm eff}\approx3$ measured in the CMB. This provides a cosmological argument excluding this class of anomaly-free currents.

\paragraph{Summary.} For exactly conserved, anomaly-free currents the gauge-invariant
completion is not excluded by the electron coupling. Either it carries a four-fermion
interaction that LEP, neutrino scattering, BBN and the CMB exclude, or it is a free
photon-coupled axion whose velocity the neutrino asymmetry set once, when the asymmetry
formed. In neither case do the cosmic neutrinos induce the signal.

\end{document}